\documentclass[a4paper]{llncs}

\usepackage{graphicx}
\usepackage{amsmath}
\usepackage{amsfonts}
\usepackage[T1]{fontenc}
\usepackage{varioref}
\usepackage{xspace}
\usepackage{paralist}
\usepackage{fancybox}
\usepackage{xcolor}
\usepackage{calc}
\usepackage{verbatim}

\usepackage{relsize}
\usepackage{booktabs}
\usepackage[scaled]{helvet}

\usepackage{pdfsetup}
\usepackage{times}

\usepackage{listingsVDM}
\usepackage{overturelanguagedef}

\definecolor{mGreen}{rgb}{0,0.6,0}
\definecolor{mGray}{rgb}{0.5,0.5,0.5}
\definecolor{mPurple}{rgb}{0.58,0,0.82}
\definecolor{backgroundColour}{rgb}{0.95,0.95,0.92}

\newcommand{\vdmkeyword}[1]{\textbf{\texttt{#1}}}

\definecolor{maroon}{rgb}{0.5,0,0}
\definecolor{darkgreen}{rgb}{0,0.5,0}
\definecolor{ao}{rgb}{0.0, 0.5, 0.0}
\definecolor{mycolor1}{rgb}{0.0, 0.53, 0.74}%
\definecolor{mycolor2}{rgb}{0.21783,0.72504,0.61926}%
\definecolor{mycolor4}{rgb}{0.93, 0.53, 0.18}%
\definecolor{plum}{rgb}{0.56, 0.27, 0.52}
\definecolor{pinegreen}{rgb}{0.0, 0.47, 0.44}
\definecolor{pthaloblue}{rgb}{0.0, 0.06, 0.54}
\definecolor{saffron}{rgb}{0.96, 0.77, 0.19}

\usepackage{cleveref}

\usepackage[caption=false]{subfig}
\usepackage{pgfplots}
\pgfplotsset{compat=newest}
\usetikzlibrary{pgfplots.groupplots, pgfplots.units}
\pgfplotsset{
	every axis label/.append style={font=\normalsize},
	tick label style={font=\small},
	/pgfplots/enlargelimits=false,
    legend style={legend pos=north east, font=\small},
    legend cell align=left,
    xlabel near ticks,
    ylabel near ticks,
	axis on top,
    highlight/.code args={#1:#2}{
        \fill [every highlight] ({axis cs:#1,0}|-{rel axis cs:0,0}) rectangle ({axis cs:#2,0}|-{rel axis cs:0,1});
    },
    /tikz/every highlight/.style={
        on layer=\pgfkeysvalueof{/pgfplots/highlight layer},
        red!10
    },
    /tikz/highlight style/.style={
        /tikz/every highlight/.append style=#1
    },
    highlight layer/.initial=axis background
}%

\lstdefinelanguage{XML}
{
  basicstyle=\ttfamily\scriptsize,
  morestring=[b]",
  moredelim=[s][\bfseries\color{maroon}]{<}{\ },
  moredelim=[s][\bfseries\color{maroon}]{</}{>},
  moredelim=[l][\bfseries\color{maroon}]{/>},
  moredelim=[l][\bfseries\color{maroon}]{>},
  morecomment=[s]{<?}{?>},
  morecomment=[s]{<!--}{-->},
  commentstyle=\color{darkgreen},
  stringstyle=\color{blue},
  identifierstyle=\color{red}
}

\usepackage{tikz}
\usepackage{balance}
\usetikzlibrary{shapes,fit,arrows,calc,arrows,arrows.meta,positioning}

\begin{document}

\title{Towards Operation Proof Obligation Generation for VDM}
\titlerunning{Towards Operation POG for VDM}

\author{Nick Battle\inst{1} \orcidID{0009-0001-1523-4964} and Peter Gorm Larsen\inst{1} \orcidID{0000-0002-4589-1500}}
\institute{Department of Electrical and Computer Engineering, Aarhus University, Denmark, \email{nick.battle@gmail.com, pgl@ece.au.dk}
}

\maketitle

\begin{abstract}
All formalisms have the ability to ensure that their models are internally consistent. Potential inconsistencies are generally highlighted by assertions called \emph{proof obligations}, and the generation of these obligations is an important role of the tools that support the method. This capability has been available for VDM tools for many years. However, support for obligation generation for explicit operation bodies has always been limited. This work describes the current state of work to address this, showing the capabilities so far and highlighting the work remaining.
\end{abstract}

\section{Introduction}
\label{sec:intro}
Tools for working with VDM-SL specifications have been able to produce \emph{proof obligations}\footnote{VDMTools~\cite{Larsen01} calls them "integrity properties".} for many years. A proof obligation (PO) is a small VDM-SL boolean expression, which should always be true -- it should be a tautology. If all obligations can be shown to be true, then the internal consistency of the specification has been verified, e.g., showing that a calculation never divides by zero.

Tool support for proof obligation generation (POG) from VDM \emph{functions} is mature, and recent work even allows POs to be checked and discharged~\cite{Battle09,FJVW13TR}. However, support for the generation of obligations from VDM \emph{explicit operations} has always been limited. This is partly due to the fact that proof rules for statements were never fully defined~\cite{Larsen95}. In addition, the depth of analysis required is greater, since operations have a more complex control flow, possibly involving loops, exceptions and side effects. In this paper, we describe the progress towards generating obligations for explicitly declared VDM operations. This is accomplished with a combination of static analysis and annotations.

Section~\ref{sec:history} looks at the background to POG in VDMTools~\cite{Larsen01}. Afterwards, Section \ref{sec:overview} gives an overview of the POG in VDMJ~\cite{Battle09,Rask2020}. Then Section \ref{sec:operation_pog} describes the new POG for operations and Section~\ref{sec:evaluation} looks at the results achieved so far. Finally, Section~\ref{sec:related} looks at related work and Section~\ref{sec:future} considers future work.

\section{History}
\label{sec:history}
The initial POG for VDM was introduced in a Masters thesis in 1997 \cite{Aichernig&97}. This was incorporated in VDMTools at IFAD, and subsequently the ideas behind it were transferred to the Overture VDM tool in 2010~\cite{Larsen&10a}. The POG was then extended with better support for recursive definitions and measures for functions later in 2010~\cite{Ribeiro&10}.

However, none of this research attempted to generate obligations for statements present inside explicitly defined operations. Originally, the VDM method was focused on refinement for operation validation~\cite{Jones90a}. This left explicit operation obligations as rudimentary, giving the basic conditions to be proved, but lacking the context needed to prove them.

The current work describes progress in providing the missing context to operation obligations, in order to allow them to be discharged.

\section{POG Overview}
\label{sec:overview}

Proof obligations are generated from VDM specifications that have passed a static type check. They are boolean expressions that express an obligation on the implementer to verify that some condition always holds. For example, a type definition with an invariant creates an obligation to make sure that at least one value passes the invariant. Similarly, within the bodies of functions and operations, obligations arise to make sure that partial operators are not applied outside their domain. For example, keys must exist in the domain of maps to which they are applied, and values must be within the size of sequences that they index.

The general form of obligations within function or operation bodies has three parts:

\begin{itemize}
    \item An outermost \emph{forall} quantifies over all possible arguments that can be passed. This includes the precondition, if any.
    \item A series of \emph{context} steps describe the path that the evaluation must take to get to the obligation point.
    \item Finally, the fundamental obligation itself.
\end{itemize}

\noindent
So in plain language, the obligation in the specification below would read "For all possible key arguments to \emph{lookup}, if the value is non-zero and it is a legal key, then the key exists in the domain of the table."

\small
\begin{vdmtt}
functions
    lookup(key:nat) r:set of nat ==
        if key <> 0 and isValid(key)
        then table(key)  -- Obligation here
        else {};

--Proof Obligation 1: (Unproved)
lookup: map apply obligation at line 4:10
(forall key:nat &
  (((key <> 0) and isValid(key)) =>
    key in set dom table))
\end{vdmtt}
\normalsize

\section{Operation POG}
\label{sec:operation_pog}

\subsection{Representing State in POs}
\label{sec:state}

One significant difference between POs from operations and functions is that the former also have access to any state data that is defined in the module. So, whereas the top-level quantifier of a function PO can consider all possible arguments passed, the quantifier of an operation has to consider all possible arguments and module states. This is represented by the \emph{Sigma} type of the model\footnote{By convention, the state of a model is called "Sigma". But in general, we mean the name of the \texttt{state} definition.}. For example:

\small
\begin{vdmtt}
state Sigma of
    sv : nat
    xv : nat
end;

operations
    op(a:nat) r:real ==
        return 1/(sv - a)  -- Obligation here
    pre sv > a;

--Proof Obligation 1: (Unproved)
op: non-zero obligation at line 8:22
(forall a:nat, mk_Sigma(sv, xv):Sigma &
  pre_op(a, mk_Sigma(sv, xv)) =>
    (sv - a) <> 0)
\end{vdmtt}
\normalsize

\noindent
Note that the \vdmkeyword{forall} quantifer includes the parameter "\texttt{a}", but it also defines variables "\texttt{sv}" and "\texttt{xv}" for the state variables, using the Sigma type and a record pattern to match the variable names.

The \texttt{pre\_op} function is subsequently called to eliminate argument/state combinations that do not match the operation's precondition. This uses a Sigma record to create a specific state vector from the variables.

This approach to state representation means that the context and the base obligation can reason about state variables by name directly, as they do in the specification itself. In VDM-SL an operation can only directly reason about the state in its own module. The position with state representation for VDM++ and VDM-RT is much more complex and is discussed in Section~\ref{sec:future}.

\subsection{Assignments to State}
\label{sec:updates}

An operation can make assignments to state variables, as well as simply reasoning about them. Therefore, the logic in an obligation that is \emph{after} a state assignment must include this change. This is achieved by using \vdmkeyword{let} expressions in the obligation context to \emph{hide} the variables bound by the outer quantifier. Thereafter, referring to "\texttt{sv}" (say) will refer to the updated definition, rather than the outer definition. For example:

\small
\begin{vdmtt}
operations
    op(a:nat) r:real ==
    (
        sv := sv + 1;
        xv := xv + sv;
        return a + 1/xv  -- Obligation here
    );
    
--Proof Obligation 1: (Unproved)
op: non-zero obligation at line 6:17
(forall a:nat, mk_Sigma(sv, xv):Sigma &
  (let sv : nat = (sv + 1) in
    (let xv : nat = (xv + sv) in
      xv <> 0)))
\end{vdmtt}
\normalsize

\noindent
This clearly works for simple variable definitions, but it also works for more complex state designators, with careful use of "\texttt{++}" and \vdmkeyword{mu} operators to create new state variable values. Note that even complex assignments only ever update one state variable (albeit with a complex new value). For example:

\small
\begin{vdmtt}
state Sigma of
    sv : seq of R
end

types
  R ::
    size : real;

op(z:nat) r:real ==
(
  sv(1).size := 456;
  return 1/len sv  -- Obligation here
);

--Proof Obligation 2: (Unproved)
op: non-zero obligation at line 12:11
(forall z:nat, mk_Sigma(sv):Sigma &
  (let sv : seq of R = sv ++ {1 |-> mu(sv(1), size |-> 456)} in
    (len sv) <> 0))
\end{vdmtt}
\normalsize

\noindent
Here, the assignment to "\texttt{sv(1).size}" only updates the "\texttt{sv}" variable, but it does so by indexing into the sequence value and then updating a record field at that position. This update is included in the obligation, using "\texttt{++}" to update the sequence with a record value that uses \vdmkeyword{mu} to modify the size field. This approach extends to arbitrary combinations of maps, sequences and records.

\subsection{Local State Handling}
\label{sec:locals}

In addition to module state, operations can define local state using assignment definitions (\vdmkeyword{dcl}) within block statements. In most respects, these are treated the same way as module state by the POG, but special care has to be taken with scoping if those state variables have been used to update module state. For example:

\small
\begin{vdmtt}
op(z:nat) r:real ==
(
    dcl a:nat := 0;
    a := a + 1;
 	 
    ( dcl b:nat := a + 1;
      sv := b );  -- NOTE: updates sv using b
    	
    ( dcl c:nat := a + 2;
      c := c + 1 );
 
    return 1/sv   --PO depends on a, b but not c
);

--Proof Obligation 1: (Unproved)
op: non-zero obligation at line 12:13
(forall z:nat, mk_Sigma(sv):Sigma &
  (let a : nat = 0 in
    (let a : nat = (a + 1) in
      (let b : nat = (a + 1) in
        (let sv : nat = b in
          sv <> 0)))))
\end{vdmtt}
\normalsize

\noindent
Here, three \vdmkeyword{dcl} definitions are created, two of them in sub-blocks. But note that the middle block uses its local "\texttt{b}" value to update "\texttt{sv}". This means that the "\texttt{b}" value must be retained at the end of its block, so that can appear in the obligation to allow the assignment to "\texttt{sv}" to be well defined. In contrast, the "\texttt{c}" value is not used to update anything and is not included in the PO.

There is considerable scope for confusion here if the specification either hides variable values or re-defines the same name in different blocks. This is discussed in Section~\ref{sec:future}.

\subsection{Alternate Paths}
\label{sec:alternates}

So far, the examples shown have only included a single control flow path from the start of the operation to the obligation point. In obligations for functions, there is only ever one path. However, for explicitly defined operations, there can be arbitrarily many paths to reach a given statement. In these cases, the separate possible paths are treated independently, generating multiple obligations for a single location.

The example below is a simple illustration. Notice that there are three obligations, corresponding to the three possible paths to reach the obligation location. Each obligation includes context that covers its particular path.

\small
\begin{vdmtt}
op(z:nat) r:real ==
(
  if z > 10 then
    if z > 100 then
      sv := 999
    else
      sv := 888
  else
    sv := z+1;

  return 1/sv     --PO#1,2,3
);

--Proof Obligation 1: (Unproved)
op: non-zero obligation at line 11:11
(forall z:nat, mk_Sigma(sv):Sigma &
  ((z > 10) =>
    ((z > 100) =>
      (let sv : nat = 999 in
        sv <> 0))))

--Proof Obligation 2: (Unproved)
op: non-zero obligation at line 11:11
(forall z:nat, mk_Sigma(sv):Sigma &
  ((z > 10) =>
    (not (z > 100) =>
      (let sv : nat = 888 in
        sv <> 0))))

--Proof Obligation 3: (Unproved)
op: non-zero obligation at line 11:11
(forall z:nat, mk_Sigma(sv):Sigma &
  (not (z > 10) =>
    (let sv : nat = (z + 1) in
      sv <> 0)))
\end{vdmtt}
\normalsize

\subsection{Ambiguous States}
\label{sec:ambiguous}

The examples given above consider direct updates to state variables by statements in an operation. However, an operation can make calls to other operations, which in turn may update the state of the system, affecting statements after the call. In general, we cannot know what an operation call will do without analyzing the entire specification. So the POG uses the concept of \emph{ambiguous states}, which record the names of variables whose actual state is not known.

A simple operation call is assumed to put every module state variable into an ambiguous state (local \vdmkeyword{dcl} state is not affected). If the operation has an \vdmkeyword{ext wr} clause that identifies named variables, then just those are marked as ambiguous, whereas if the operation is \vdmkeyword{pure} is it known not to update anything. The return value from an operation is generally unknown, so any variable assigned a value using the result from an operation call is marked as ambiguous.

If any variables in the context of an obligation are ambiguous, the PO is produced but marked as "Unchecked".

Note that ambiguous variables can subsequently be disambiguated by being assigned with values that are calculated from unambiguous values.

\subsection{Atomic Updates}
\label{sec:atomics}

The \vdmkeyword{atomic} statement in operations has to have special handling because of its semantics. These statements define a collection of assignments which all happen \emph{atomically} -- that is, the effect of each assignment is not visible to the others. In practice, this is equivalent to calculating the RHS values for them all, then making the assignments with any state invariant disabled, and finally checking the invariant after all the assignments have completed. This process is therefore reflected in the obligation context for atomic statements. For example:

\small
\begin{vdmtt}
state Sigma of
    sv : real
    xv : real
inv s == s.sv <> s.xv
end

op(a:nat) ==
    atomic ( sv := xv; xv := sv );  -- Obligation here

--Proof Obligation 1: (Unproved)
(forall a:nat, mk_Sigma(sv, xv):Sigma &
  (let $atomic1 : real = xv in
    (let $atomic2 : real = sv in
      (let sv : real = $atomic1 in
        (let xv : real = $atomic2 in
          let s = mk_Sigma!(sv, xv) in ((s.sv) <> (s.xv)))))))
\end{vdmtt}
\normalsize

\noindent
Note the use of the maximal operator in the final \texttt{mk\_Sigma!}. This allows the state record to be created without the invariant, for the purpose of explicitly checking the invariant on the resulting value.

\subsection{Post-conditions}
\label{sec:post}

Post-conditions in operations are able to reason about the "old" value of state variables, using a tilde syntax, like \texttt{var\~}. This poses a problem for the obligation generator, partly because the original value of variables must be represented, and partly because the tilde syntax can only be used legally in a post-condition clause.

Old variables and return values are therefore handled as follows:

\small
\begin{vdmtt}
state Sigma of
    sv : nat
end

op(z:nat) r:real ==
(
  sv := z;
  sv := sv * 2;
  return sv + 1
)
post r > 0 and sv > sv~;

-- Proof Obligation 1: (Unproved)
(forall z:nat, mk_Sigma(sv):Sigma &
  (let sv$ = sv in             <-- Old state
    (let sv : nat = z in
      (let sv : nat = (sv * 2) in
        (let r = (sv + 1) in   <-- return sets "r"
          ((r > 0) and (sv > sv$)))))))
\end{vdmtt}
\normalsize

\noindent
If an operation has a post-condition which reasons about the original value of state variables, these are captured at the start of the obligation, using a substitution for "\$" in place of the tilde. Return statements create a context that defines the returning variable, or "\vdmkeyword{RESULT}" if none is defined. This can then be used in the postcondition assertion at the end.

\subsection{Loop Invariants}
\label{sec:loops}

The analysis of loop statements requires a \emph{loop invariant} to be specified for each loop. Proof obligations can then be generated for points before and during the loop, and the invariant also appears in the context of obligations after the loop.

Loop invariants are defined using a \texttt{@LoopInvariant} annotation \cite{Battle25}, which takes a single expression as an argument. The expression is obliged to reason about the relationship between all of the variables that the loop modifies. It is then used in obligations.

In the example below, we use \emph{inline functions} within the obligation to allow the loop to be represented as a recursive function. The \texttt{body} performs the changes to the updatable variables in the loop; the \texttt{invariant} is the expression from the annotation; and the \texttt{loop} checks the while condition, then checks the invariant before and after the updates, before recursing to process the next loop iteration.

This method of creating loop obligations should be regarded as experimental and the tools do not currently produce the PO shown\footnote{Currently, valid obligations are produced for the point before the loop, and at the start and end of the first loop iteration.}. However, we believe that this is a possible route to handling loop obligations.

\newpage
\small
\begin{vdmtt}
state Sigma of
  s : seq of int
end

op(data:seq of int) ==
(
  dcl count : int := 0;
  s := data;

  -- @LoopInvariant(count + len s = len data);
  while s <> [] do
  (
    s := tl s;
    count := count + 1
  )

  -- Here, invariant holds and s = []
  ...
);

--Proof Obligation 1: (Unproved)
(forall data:seq of int, mk_Sigma(s):Sigma &
    let body: seq of int * int +> seq of int * int
        body(s, count) ==
            (let s : seq of int = (tl s) in
                (let count : int = (count + 1) in
                    mk_(s, count))),

        invariant: seq of int * int * seq of int +> bool
        invariant(s, count, data) == 
            ((count + (len s)) = (len data)),
            
        loop: seq of int * int * seq of int +> bool
        loop(s, count, data) ==
            s <> [] =>
                invariant(s, count, data) and
                let mk_(s, count) = body(s, count) in
                    invariant(s, count, data)
                    and loop(s, count, data)
    in
        (let count : int = 0 in
            (let s : seq of int = data in
                (loop(s, count, data)))))

\end{vdmtt}
\normalsize
\noindent
The example above is a \vdmkeyword{while} loop, but a similar approach works for \vdmkeyword{for} loops if the \texttt{loop} function is passed the variable(s) to modify and test as the loop progresses.

If the annotation is missing from a loop\footnote{Obviously they are missing from historical specifications.}, the POG cannot produce the corresponding obligations. In this case, the variables updated by the loop are marked as ambiguous and any obligations resulting from the body of the loop are marked as "Unchecked".

Note that the type-checker verifies that the \texttt{@LoopInvariant} expression reasons about all the modified variables in the loop. The invariant is also checked by the interpreter runtime, so ad-hoc testing or combinatorial testing may also pick up failures.

\subsection{Obligation Correctness}
\label{sec:correctness}

Obligations are intended for discharge by the proof process, but the development of the obligation generator \emph{itself} must be validated to ensure that the POs produced are sufficient and correctly describe the obligations placed on the specifier.

This process is hampered by the lack of a proof theory for explicit operations in \cite{Larsen95}. But the general approach we adopt is to break down the flow of control in operations to give a set of possible paths. Relevant actions that the operation makes along a path are represented by clauses in the PO context, as described earlier. If the sequential composition of these clauses always results in a semantic equivalence to the operation itself, then the complete obligation will correctly represent the state of the system on that one path.

This approach seems promising for the cases considered so far, as demonstrated in this section. However, it may not be sufficient for more complex cases (see Section \ref{sec:future}). The authors hope that the technique of using in-line functions (see \ref{sec:loops}) can be expanded to deal with more complex cases.

\section{Preliminary Evaluation}
\label{sec:evaluation}

The new POG has been tested on the VDM-SL example suite that comes with Overture~\cite{Coleman&12b}. The 50 example specifications generate 5500+ obligations. These are still being checked, but the POG does not crash, and tools such as QuickCheck~\cite{Battle2024} can attempt to discharge the obligations generated.

Before the changes described in this work, about 21\% of the obligations generated by the example suite were marked as "Unchecked"; with the new POG, about 9.6\% of them remain "Unchecked". This indicates that, although the techniques deployed so far do not tackle complex cases, they still result in a significant increase in the number of obligations that can be checked by proof tools.

With the new POG, around 14\% (811) of the obligations are \emph{failed} by QuickCheck. Most failures are cases where operations are missing obvious constraints. But in at least one case, the new POG caught a very subtle error in the "\texttt{TicTacToe}" model, where the "move" operation did not have a sufficiently strong precondition. These problems were previously missed because the corresponding obligations were incomplete and marked "Unchecked".

\section{Related Work}
\label{sec:related}

The process of calculating proof obligations for imperative VDM operations is similar to other formal languages, most notably \emph{Dafny}~\cite{leino2010dafny}. Some of the work here was guided by Dafny's \emph{Verification Conditions} (VCs).

\section{Future Work}
\label{sec:future}

The work outlined above is not complete. The following areas or issues remain to be completed or solved:

\begin{itemize}
    \item Loop invariants are the right direction for dealing with loops, but the generated POs are currently insufficient. For example, although the simple example in Section~\ref{sec:loops} works, it does not account for obligations within the loop body or the effect of loops on subsequent obligations on the path. We also probably need a \texttt{@LoopTermination} annotation, which would be similar to a recursive measure, raising POs to verify that each loop gets closer to termination.
    \vspace{10pt}
    \item Statements that handle exceptions (\vdmkeyword{always}, \vdmkeyword{trap} and \vdmkeyword{tixe}) cause control flows that are currently too complex to handle. Every statement within the body of these statements that could raise an exception effectively creates a set of new paths, one for each type of exception that could be thrown. Since VDM operations do not explicitly declare the exceptions that they can raise (i.e. there is no equivalent of Java's "throws" clause), this requires a deep analysis of the specification, and probably requires a pessimistic assumption about whether a given exception can actually occur.
    \vspace{10pt}
    \item Specifications that include variable hiding can easily confuse the POG and produce invalid POs. Currently, a few cases of variable hiding are explicitly checked by the POG and POs are subsequently marked "Unchecked". But this approach is not generalized. A more robust approach would understand the variable hiding in the specification and perhaps rename variables in the obligation to compensate. The best advice is to avoid variable hiding in specifications and respect the type checker warnings if you get them.
    \vspace{10pt}
    \item A deeper analysis of a specification's call graph and variable assignments may allow a more sophisticated technique than marking state as ambiguous after operation calls. Unfortunately, analyzing the call graph can become uncertain, given that we cannot predict which exceptions will be thrown. So such analysis would have to be pessimistic and \emph{assume} that any path that could modify the state would actually do so. The result may produce obligations for paths that are actually not reachable.
    \vspace{10pt}
    \item As mentioned in Section \ref{sec:correctness}, the correct operation of the POG itself must be determined. This should ultimately be linked to a proof theory for VDM operations.
    \vspace{0pt}
    \item VDM++ and VDM-RT cause a host of problems that have yet to be solved. Unlike the \emph{mk\_Sigma} record of VDM-SL, the state of these dialects cannot be easily represented, or passed to precondition functions. Furthermore, the state includes all accessible \vdmkeyword{static} fields in the model. VDM++ models can include multiple threads with synchronization clauses, which massively complicates the POG analysis. And VDM-RT introduces distributed behaviour with CPUs over connecting busses, which is even more complicated.
\end{itemize}

\vspace{10pt}
\paragraph{\textbf{Acknowledgements}}
We are grateful for the support of the European Union, Aarhus University, Newcastle University and the Grundfos Foundation.
We also thank the reviewers for their valuable feedback on the original version of this paper.

\bibliographystyle{splncs04}
\bibliography{au.bib}

\end{document}